# Compiler-Assisted Workload Consolidation For Efficient Dynamic Parallelism on GPU

Hancheng Wu*, Da Li*, Michela Becchi

Dept. of Electrical and Computer Engineering
University of Missouri
Columbia, MO, USA
{hancheng.wu, da.li}@mail.missouri.edu, becchim@missouri.edu

*Abstract*— GPUs have been widely used to accelerate computations exhibiting simple patterns of parallelism – such as flat or two-level parallelism – and a degree of parallelism that can be statically determined based on the size of the input dataset. However, the effective use of GPUs for algorithms exhibiting complex patterns of parallelism, possibly known only at runtime, is still an open problem. Recently, Nvidia has introduced Dynamic Parallelism (DP) in its GPUs. By making it possible to launch kernels directly from GPU threads, this feature enables nested parallelism at runtime. However, the effective use of DP must still be understood: a naïve use of this feature may suffer from significant runtime overhead and lead to GPU underutilization, resulting in poor performance.

In this work, we target this problem. First, we demonstrate how a naïve use of DP can result in poor performance. Second, we propose three workload consolidation schemes to improve performance and hardware utilization of DP-based codes, and we implement these code transformations in a directive-based compiler. Finally, we evaluate our framework on two categories of applications: algorithms including irregular loops and algorithms exhibiting parallel recursion. Our experiments show that our approach significantly reduces runtime overhead and improves GPU utilization, leading to speedup factors from 90x to 3300x over basic DP-based solutions and speedups from 2x to 6x over flat implementations.

## I. Introduction

Graphics processing units (GPUs) have proven to successfully accelerate regularly structured, data intensive applications. Applications that naturally map onto the GPU hardware typically exhibit simple parallelism patterns – such as flat and two-level parallelism – and a degree of parallelism that can be determined prior to kernel launch based on the size of the inputs. However, many common applications operate on irregularly structured data (such as graphs and meshes) and present data-dependent control flows and irregular memory access patterns, resulting in unpredictable runtime behaviors. Examples of such applications include adaptive meshes, recursive algorithms and computations with irregular loops [1]. The use of flat parallelism to implement these applications can lead to uneven work distribution across threads and, ultimately, to inefficient codes. In fact, in the presence of flat kernels, fine-grained partitioning of the work across threads may lead to a large number of idle or lightly utilized threads, while coarse-grained partitioning of the work may cause uneven work distribution among threads. On the other hand, the use of more adaptive parallelism patterns, such as nested parallelism, can better capture the dynamic nature of the workload within unstructured applications.

Since the Kepler architecture, Nvidia has introduced in its GPUs a new feature, called *dynamic parallelism* (DP), which makes it possible for GPU threads to dynamically spawn GPU kernels. Dynamic parallelism has also been recently added to the OpenCL 2.0 standard. By supporting nested kernel invocations, this feature enables nested parallelism on GPU, and can facilitate dynamic load balancing, data-dependent execution and the parallelization of recursive algorithms. However, the effective use of DP is not a trivial matter. Basic implementations of adaptive kernels that spawn child kernels on a per-thread basis whenever new work is locally generated tend to perform a large number of small kernel launches. It has been shown that, due to the runtime overhead associated with nested kernel calls, these implementations often lead to significantly degraded performances, even worse than those of flat parallel variants of the same algorithms [2-4].

In this work, we address this problem and propose a workload consolidation approach to improve the performance of applications relying on DP. Specifically, we consolidate into a single nested kernel the workload belonging to kernels that would be spawned by multiple GPU threads. We consider performing kernel consolidation at three granularities: *warp-*, *block-* and *grid-level*, whereby the consolidation involves the kernels launched by all the threads within a warp, all the threads within a block or the entire grid, respectively. We evaluate our consolidation mechanisms on applications exhibiting two computational patterns: parallel irregular loops and parallel recursion.

In summary, we make the following contributions:
- We propose a software-based workload consolidation mechanism for kernels relying on DP.
- We integrate our consolidation schemes in a directive-based compiler. By automating our code transformations, we allow programmers to write simple code focusing on functionality rather than on performance.
- We observe that static methods to configure the degree of multithreading of GPU kernels (e.g. CUDA occupancy calculator) are ineffective in the presence of DP. We propose a systematic way to configure dynamic kernel launches.
- We evaluate our proposal on seven applications that include irregular loops and parallel recursion, and provide insights on the effective utilization of DP. Our results show that our code transformations significantly reduce runtime overhead and improve GPU utilization, leading to speedup factors from 90x to 3300x over basic DP-based codes and speedups from 2x to 6x over flat implementations.

---

\* = authors contributed equally

## II. BACKGROUND

### A. Dynamic Parallelism

Traditionally, only CPU threads can launch GPU kernels. *Dynamic Parallelism*, a feature added to OpenCL 2.0 standard and supported by Nvidia GPUs with compute capability 3.5 and above, makes it possible for GPU threads to launch GPU kernels. Kernel launches can be nested and the deepest nesting level supported is currently 24. Kernels launched from different blocks or streams can execute concurrently, and up to 32 concurrent kernels are currently allowed on Nvidia GPUs. A parent kernel will return only after all its child kernels have completed; however, the order of their execution is unknown unless these kernels are explicitly synchronized by a *cudaDeviceSynchronize* call. For each nesting level up to an explicit synchronization, parent kernels may be temporarily swapped out to free up GPU resources and allow the execution of their child kernels. Pending kernels due to either unresolved dependencies or lack of available hardware resources are fed to a temporary buffer. Global memory data are visible to both parent and child kernels, while shared and local memory variables are visible only within the kernel where they are declared.

### B. Application Characterization

We consider two computational patterns that can benefit from DP: irregular loops and parallel recursion.

**Irregular Loops** - Irregular loops are characterized by an uneven work distribution across loop iterations. For example, nested loops where the number of iterations of inner loops varies across the iterations of outer loops are irregular. These loops can be found in many applications, such as sparse matrix operations and graph traversal algorithms that rely on the commonly used Compressed Sparse Row representation of matrix and graph data structures [5]. The degree of parallelism within irregular loops is typically data dependent and known only at runtime. Due to their nature, the flat parallelization of these loops can cause work imbalance across threads, possibly leading to GPU underutilization and limited performance. For example, if loop iterations are distributed equally to threads, the unbalanced work distribution across iterations (and threads) will lead to warp divergence. Irregular loops can benefit from the use of DP for load balancing [6]. Specifically, overloaded threads can spawn child kernels and assign (part of) their work to them, thus limiting warp divergence.

**Parallel recursion** - Nested parallelism also arises in the presence of parallel recursion. While some recursive algorithms can be made iterative through various code transformation techniques (e.g. tail-recursion elimination, auto-ropes and other flattening techniques [7-9]) and subsequently undergo flat parallelization, recursion cannot be always eliminated. Before the introduction of DP on GPU, parallel recursion required both CPU and GPU intervention. Specifically, the CPU would control the flow of recursion and initiate the recursive calls, whose execution could then be offloaded to the GPU in the form of parallel kernels. This approach requires one kernel launch for every recursive call and incurs high CPU-GPU communication overhead. By enabling kernel launches from GPU threads, DP allows the recursive control flow to reside directly on the GPU. This, in turn, allows reducing the CPU-GPU communication and the kernel launch overhead. The most natural way to implement a parallel recursive function on GPU is by directly porting to this platform a CPU parallelization of that function and allowing each GPU thread to spawn a recursive kernel whenever the CPU code would perform a recursive call. As we will show, GPU implementations following this pattern often perform a large number of small kernel invocations, leading to substantial kernel launch overhead and hardware underutilization. Our proposed consolidation schemes target this problem.

## III. MOTIVATION

In this section, we first present the basic use cases of dynamic parallelism. Then we show how inefficient the basic implementation is and explain why using DP can lead to performance degradation. These motivate us to propose our compiler-assisted workload consolidation solution.

### A. Basic DP-Code Template

Figure 1(a) shows a basic code template for parallel kernels that use DP [2]. As in all GPU kernels, each thread (or thread-block) is assigned some data to process (or *work items*). Each thread (thread-block) initially performs some work (*prework*) on the data assigned to it. Then, depending on the outcome of a condition, the thread (thread-block) either spawns a child kernel to execute some more *work*, or performs that *work* on its own. Finally, the thread (thread-block) may optionally perform additional work (*postwork*). Note that both irregular loops and recursive algorithms fit in this code template. The only difference between these two computational patterns is the following: in irregular loops parent and child kernels are different, and the child kernel is generally used for load balancing purposes; in parallel recursion, parent and child kernels are identical. Figures 1(b) and (c) illustrate this basic code template on two algorithms: one with an irregular loop (*single source shortest path*), and the other with parallel recursion (*recursive tree traversal*).

In the *SSSP* kernel, each GPU thread processes the neighbors of an assigned node. Because the number of neighbors may vary from node to node, the workload may be unevenly distributed across threads. To address this problem, each thread checks whether the amount of work assigned to it (*neighbors.size*) is larger than a given threshold. If this is the case, it spawns a child kernel and delegates the work to it; otherwise it performs the work on its own. Because the GPU hardware schedules different kernels independently, this mechanism allows redistributing the work to the GPU resources, potentially leading to better GPU utilization.

In the tree traversal kernel, each thread is assigned a *child* of a given node. Initially, the thread checks whether *child* has no children. In this (base) case, the thread performs *leafnode_work*; otherwise, it spawns a kernel recursively and delegates to it the processing of its assigned node.

```
__global__ void parent_kernel() {
    work_item= get_work_item(…)
    prework(work_item)
    if (condition)
        child_kernel<<<block_dim, thread_dim>>>(…, work_item, …)
    else
        work(work_item)
    postwork(work_item)
}
```
(a) Basic code template for kernels using dynamic parallelism

```
__global__ void sssp() {
    int i = blockDim.x * blockIdx.x + threadIdx.x
    neighbors = get_neighbors(i)
    if (neighbors.size>THRESHOLD)
        process_neighbors<<<block_num, thread_num>>>(neighbors)
    else {
        work(neighbors)
    }
    postwork()
}
```
(b) Basic implementation of SSSP with dynamic parallelism

```
__global__ void tree_traversal( node_t node) {
    int i = threadIdx.x
    node_t child = node.children[i]
    if (child.children.size>0)
        tree_traversal<<<1, child.children.size>>>(child)
    else {
        leafnode_work()
    }
    postwork()
}
```
(c) Basic implementation of tree traversal with dynamic parallelism

Figure 1. Basic-dp code template and sample codes

The examples above illustrate "basic" implementations of irregular loops and parallel recursion that rely on DP. For irregular loops, DP is used to redistribute unbalanced work: in this case, the thread (thread-block) executing an iteration of the loop invokes a nested kernel to offload the excess work to it. In case of parallel recursion, this basic code variant results from simply porting the CPU implementation of a parallel recursive function to GPU. Although more complex implementations are possible, these basic code variants require minimal effort from the programmer. However, as discussed below, this basic use of DP can incur significant overhead and lead to poor performance.

## B. Limitations of Dynamic Parallelism

The effectiveness of DP is affected by different factors: sources of runtime overhead and GPU utilization. Below, we detail each of these aspects.

**Kernel Launch Overhead** - To launch a kernel, the CUDA driver and runtime need to parse the parameters list, buffer the values of these parameters, and dispatch the kernel. These steps have an associated overhead. This overhead is negligible when the number of nested kernels is small, but can accumulate and become significant in the presence of numerous kernel launches [2, 3].

**Kernel Buffering Overhead** - Kernels waiting to execute are inserted in a pending buffer. Since CUDA 6, this buffer consists of two pools: a fixed-size pool and a variable-size virtualized pool. The fixed-size pool incurs lower management overhead but by default can only accommodate a maximum of 2048 pending kernels. When it becomes full, pending kernels are fed to the virtual pool, which incurs extra management overhead. Applications spawning a large number of nested kernels can exhaust the fixed-size pool and experience performance degradation due to the virtual pool's overhead. It is possible to increase the capacity of the fixed-size pool through the CUDA function *cudaDeviceSetLimit*. However, this will result in a higher global memory reservation.

**Synchronization Overhead** - If there exists explicit synchronization between parent and child kernels, in order to free resources for the execution of child kernels, parent kernels will be swapped out into global memory. For each level up to the maximum level where they synchronize, up to 150 MB memory may be reserved for swapping. These extra memory transactions are source of additional overhead.

**Effect of the kernel configuration** – It is well known that the full use of the GPU hardware requires massive multithreading. CUDA currently allows up to 32 kernels to execute concurrently on a GPU. However, if configured to use small thread configurations, even 32 concurrent kernels may underutilize the GPU. Meanwhile, there is a limit on the maximum number of blocks that can be concurrently active. Thus, configuring nested kernels with a large number of blocks will limit the number of kernels executing in parallel. As a result, it is important for programmers to carefully select thread configurations that allow both good device utilization and desired level of concurrency.

The kernel launch, buffering and synchronization overheads can be reduced by limiting the number of kernel launches performed. In addition, to avoid GPU underutilization, it is important to avoid small kernel configurations that would lead to low occupancy even in the presence of kernel concurrency. In general, DP codes resulting in a large number of small kernel invocations tend to experience poor performance. In our previous work [3], we have shown that, due to the large number of small kernel calls they invoke, DP codes following the basic template in Figure 1 can underperform flat implementations of the same algorithms by up to a factor of 1000.

## IV. METHODOLOGY

In this section, we present our workload consolidation mechanism designed to avoid the performance degradation associated with the basic use of dynamic parallelism described in Section III.

## A. Workload Consolidation

The idea at the basis of workload consolidation is fairly simple: by aggregating kernels spawned by different threads into a single or few consolidated kernels, it is possible both to decrease the number of nested kernel invocations, thus limiting DP overhead, and to increase the degree of multithreading of the nested kernels invoked, thus increasing their GPU utilization. In order to perform workload consolidation, we buffer the work associated to the kernels to be consolidated, and we defer the handling of this aggregated work to the launch of one or more child kernels. Since in the SIMT model all threads execute the same instructions on

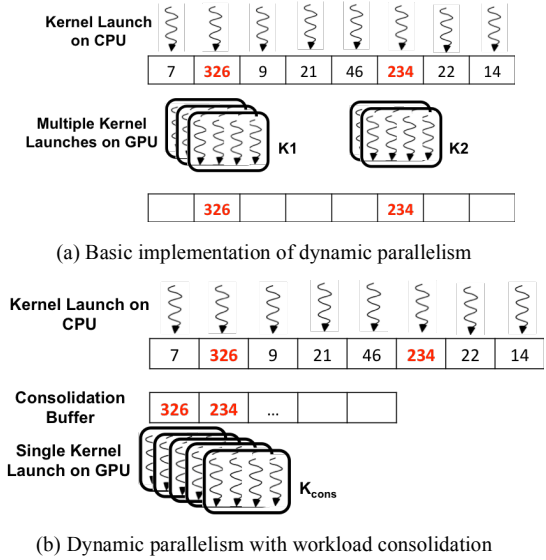

(a) Basic implementation of dynamic parallelism

(b) Dynamic parallelism with workload consolidation

Figure 2. Workload consolidation – illustration

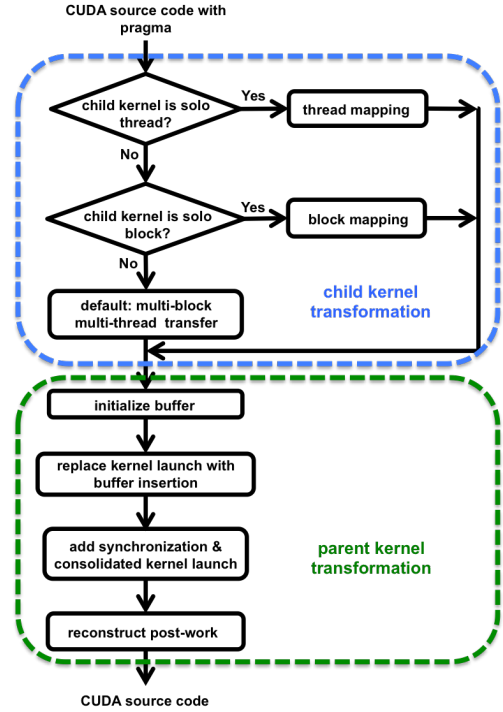

Figure 3. Kernel transformation flow

different data, in order for a thread to buffer work, it will be sufficient for the thread to buffer the pointer(s) or index(es) to the data to be processed. This method requires barrier synchronization between the buffer insertions and the consolidated kernel launches.

This high level idea is illustrated in Figure 2. In the figure, the numbers in the array indicate the amount of work assigned to each thread, and the numbers in red indicate large workloads that need to be redistributed through the launch of nested kernels. In Figure 2(a), two child kernels ($K_1$ and $K_2$) are invoked: one to process 326 and the other to process 234 work items. Workload consolidation, illustrated in Figure 2(b), first inserts the work associated to $K_1$ and $K_2$ into a consolidation buffer. It then launches a single child kernel ($K_{cons}$) to process all the work in the buffer. $K_{cons}$ will have a larger thread configuration than $K_1$ and $K_2$.

### B. Consolidation Granularity

CUDA programming model has four levels of parallelism: thread, warp, block and grid. While threads, blocks and grids are exposed to programmers, warps are implicitly defined as groups of 32 threads that execute in lockstep and have contiguous identifiers. In the DP execution model, kernel launches are performed by threads. We consider three consolidation granularities: warp-, block- and grid-level.

**Warp-level consolidation** uses a buffer to aggregate work from the threads within a warp and launches one kernel per warp. This consolidation method can reduce the number of kernel invocations at most by a factor of 32. The benefit of warp-level consolidation is that the synchronization overhead is minimized because no additional synchronization is required beside the implicit one due to SIMD execution.

**Block-level consolidation** aggregates work associated to threads within a block and launches a kernel per block. This method can reduce the number of kernel invocations beyond what allowed by warp-level consolidation. However, this consolidation scheme requires a block-level synchronization (__syncthreads) after the buffer insertions, leading to higher synchronization overhead than the warp-level variant.

**Grid-level consolidation** aggregates work from all threads in a grid and then launches a single child kernel. Since CUDA does not provide global synchronization within a kernel, this consolidation method requires a customized barrier synchronization (we will discuss this aspect later). Because of this, grid-level consolidation suffers from the highest synchronization overhead.

### C. Kernel Transformations

The overall kernel transformation flow is shown in Figure 3. The input is DP-based CUDA code annotated with the described pragma directive, and the output is the consolidated CUDA code. The kernel transformation process consists of two steps: (1) child kernel and (2) parent kernel transformation. For irregular loops, parent and child kernels are different, and the two code transformations are applied separately to each. For recursive computations, parent and child kernels are the same, and the two transformation steps are applied to the single input kernel sequentially.

**Child kernel transformation** – This phase transforms the input child kernel into a consolidated child kernel. The new kernel fetches work from the consolidation buffer and processes that work according to the code in the input child kernel. Whenever possible, we generate moldable kernels [10] (that is, kernels with tunable kernel configuration). The way the original code is mapped to threads and blocks in the consolidated kernel depends on the configuration of the input child kernel. Specifically, we identify the following cases:

*Solo thread*: The input child kernel consists of a single block and a single thread (e.g. quick sort in CUDA SDK). In the consolidated child kernel, each thread will fetch a work

Table I. Clauses of our workload consolidation compiler directive

| Clause | Argument | Description | Optional |
|---|---|---|---|
| consldt | granularity: *warp*, *block*, *grid* | Workload consolidation granularity | No |
| buffer | type: *default*, *halloc*, *custom* | Buffer allocation mechanism | Yes |
| | perBufferSize: integer or variable name | Buffer size | |
| | totalSize: integer | Total size of all buffers | |
| work | varlist: list of indexes or pointers to work | List of variables to be stored in buffer | No |
| threads | thread number: integer | Number of thread/block for consolidated kernel | Yes |
| blocks | block number: integer | Number of blocks for consolidated kernel | Yes |

```
__global__ void parent_kernel() {
  work_item = get_work_item(…)
  prework(work_item)
  if (condition) {
    #pragma dp consldt(block) buffer(default, 256) work(work_item)
    child_kernel<<<block_dim, thread_dim>>>(…, work_item, …)
  } else  work(work_item)
  postwork(work_item)
}
```
(a) Annotated CUDA code (parent kernel)

```
__global__ void parent_kernel() {
  work_item = get_work_item(…);
  prework(work_item)
  if (condition)  insert_buffer(curr)
  else  work(work_item)
  synchronize
  if (thread_id==selected)
    child_kernel_consolidate<<<b_dim_con, t_dim_con>>>()
  synchronize
  postwork(work_item)
}
```
(b) Generated CUDA code (parent kernel)

Figure 4. Example of use of our workload consolidation compiler directive: (a) original annotated code and (b) generated CUDA code. The generated CUDA code is generic (it applies to all consolidation schemes). For block-level consolidation, the *synchronize* primitive used is *__syncthreads*. For warp-level consolidation, the synchronization is implicit and no synchronization primitive is required. For grid-level consolidation, *synchronize* is a custom global synchronization primitive.

item (if available) from the consolidation buffer, and it will process that work item exactly as the original kernel does. To make the kernel moldable, we allow threads to fetch work from the buffer repeatedly until the buffer becomes empty.

*Solo block:* The input child kernel consists of a single block with multiple threads, and these threads operate cooperatively. In the consolidated child kernel, each block will fetch a work item (if available) from the consolidation buffer, and the threads within the block will cooperatively process that work item as in the original kernel. To make the kernel moldable, we allow blocks to fetch work from the buffer repeatedly until the buffer becomes empty. If the original child kernel is moldable, the number of threads per block in the generated child kernel will also be tunable; else, the two kernels will have the same block size.

*Multi-block* and *multi-thread:* When the original child kernel uses multiple blocks and threads per block, each work item is processed by all threads cooperatively. In this case, in the transformed kernel we use a for-loop to wrap the code of the original child kernel. The generated kernel will extract work from the buffer iteratively, and all threads will process cooperatively each work item. In this case, the transformed kernel is moldable only if the original kernel is such.

**Parent kernel transformation** – We divide the parent kernel into three sections: *prework*, *child kernel launch*, and *postwork* (Figure 1(a)). The prework and postwork represent the processing done before and after the child kernel launch, respectively. Many kernels do not include any postwork. The code transformations required in the parent kernel are: (1) buffer allocation (before prework); (2) prework insertion; (3) replacement of the child kernel launch with buffer insertions; (4) insertion of the required barrier synchronization primitive; and (5) postwork transformation. As shown in Figure 3, steps (1)-(3) are fairly mechanic; however, steps (4) and (5) require some consideration.

If the original kernel includes barrier synchronization between the child kernel launch and the postwork, such synchronization must be preserved in the consolidated parent kernel. For warp-level consolidation, this problem is automatically solved by the implicit barrier synchronization due to the lockstep execution of the threads within a warp. For block-level consolidation, CUDA provides a block-level barrier synchronization primitive (*__syncthreads*). Grid-level consolidation requires more thought. First, the only global barrier synchronization provided by CUDA is the implicit one at the end of a kernel launch. However, using this mechanism would require splitting the parent kernel into two: a kernel handling the prework and the child kernel launch, and a kernel handling the postwork. In addition, CPU intervention would be required to invoke the postwork-kernel. This would be problematic in case of recursion, as it would require returning the control to the CPU after each child kernel launch, and to have calls to the postwork-kernel stacked on CPU. In other words, the CPU would acquire full recursion control, leading to the overheads discussed in Section II.B. To address this problem, we implement a custom global synchronization mechanism that can be invoked from the GPU (see Section E.2). Second, a global synchronization may cause a deadlock when active blocks on GPU are suspended at the global barrier while pending blocks are waiting for active blocks to finish. To address this problem, we consolidate the postwork into a single kernel. The last block to complete its buffer insertions will then launch the consolidated child kernel, wait for its completion and then launch the consolidated postwork kernel. The other blocks will simply exit after completing their buffer insertions. Finally, the required barrier synchronization between the child kernel launch and the postwork is handled by inserting a *cudaDeviceSynchronize* call between the invocations of these two consolidated kernels. Dependencies between the prework and the postwork are handled by duplicating in the postwork the relevant portions of prework.

### D. Compiler Directive Design

In order to direct the code transformations performed by the compiler, we provide a single directive that can be applied to generic DP-based code following the template in Figure 1. This directive allows identifying the child kernels to be consolidated and the work to be buffered. The grammar of the directive is: **'#pragma dp [clause+]'**. Table 1 lists the clauses available, which specify the consolidation granularity, the type and size of the consolidation buffer, the indexes or pointers to the work items to be buffered and the configuration of the consolidated kernel. Some of these clauses are optional and programmers can use them for further optimization and performance tuning. For instance, developers can optionally specify the configuration of the consolidation buffers and the one of the child kernels. We provide more details on these options in Section IV.E.

Figure 4(a) illustrates the use of our proposed compiler directive to annotate the original CUDA code. In this case, block-level consolidation is selected, the buffer can have at most 256 elements and is instantiated with the default CUDA allocator, and variable *curr* is buffered. The generated code is shown in Figure 4(b) (in this particular case, the synchronization primitive in use is *__synchthread*).

*E. Implementation Details*

This section describes the implementation details of our source-to-source compiler, which converts annotated CUDA code into consolidated GPU kernels. Our compiler is implemented using the ROSE compiler infrastructure [11].

**Rose compiler infrastructure** – ROSE (version 0.96.a) incorporates Edison Design Group (EDG) Frontend 4.0 that supports the parsing of CUDA, C and C++ code. We use its parser building APIs to implement the parser for the *pragma* directive. The *pragma* information is linked to the abstract syntax tree (AST). Based on the directive information and the AST, we customize the traversal and transformation functions to generate a new AST, which is then fed to and unparsed by the backend of ROSE to generate the consolidated parent and child kernels.

**Consolidation Buffers** – The design of the consolidation buffers involves several considerations, some of them leading to the need for the directive clauses listed in Table I.

*Memory selection*: The consolidation buffer can be either implemented in global or in shared memory. Global memory provides slower access but is visible to both parent and child kernels. Conversely, shared memory is faster but private to each block (and thus not visible within nested kernels). While parent kernels could fill temporary buffers in shared memory and then copy them into global memory to allow access by child kernels, the limited size of shared memory makes this solution not scalable. As a result, we store the consolidation buffers solely in global memory.

*Dynamic allocation method:* For the allocation of the consolidation buffers, we allow three alternatives: (1) the default allocator provided by CUDA; (2) the open-source *halloc* memory allocator for GPU [12]; and (3) a customized allocator that leverages a pre-allocated memory pool.

*Buffer size for customized allocator:* Due to the irregular nature of nested parallelism, the buffers required by different consolidated kernels may vary in size. When using the pre-allocated memory pool, the programmer needs to set both its size and the size of the portion of the memory pool allocated to each buffer (recall that in warp/block-level consolidation every warp/block uses a consolidation buffer). The size of the pre-allocated memory pool (500MB by default) can be specified using the *totalSize* argument in the *pragma* directive. The per-buffer size (*perBufferSize*) is predicted as:

$$totalThread * totalBuffVar * const$$

where *totalThread* is the total number of threads from which we consolidate the child kernels, *totalBuffVar* is the number of buffered variables (indexes or pointers) per work item; and *const* is a constant (default value: 4) that estimates the number of work items assigned to a single thread. We have observed that, in most cases, the *perBufferSize* can be determined from a runtime variable that indicates a property of the current work item. For instance, for the tree applications in our benchmarks, the buffer size can be derived from the variable that indicates the number of children of a given node. If the user cannot provide such variable, a constant may also be specified to its best estimation. Our customized allocator can utilize the information from the *#pragma* to allocate properly sized buffers from the pre-allocated global memory pool for different consolidation granularities. Notice that for grid level consolidation, only one buffer is required for each grid: in this case, the grid can directly utilize the whole memory pool and the *perBufferSize* is ignored.

**Global Barrier Synchronization on GPU** – The global barrier synchronization is implemented using a counter whose value is initialized to the number of blocks executed. When a block reaches the barrier, the first thread in the block decrements the counter by one atomically. A counter decrement to zero indicates that the last block has reached the barrier.

**Kernel Configuration Handling** – When launching kernel calls, it is common wisdom to select a configuration that achieves high device occupancy, which is defined as the ratio of the number of active warps to the number of maximum active warps that the device can host. Although higher occupancy does not always guarantee higher performance, it usually produces a good enough result that can be further tuned. The use of the *CUDA Occupancy Calculator* allows finding a kernel configuration ($B$, $T$) that maximizes the occupancy for a single kernel. However, concurrent kernels must share the GPU resources, and thus such configuration will not be optimal for concurrent kernels initiated with DP [10]. To allow multiple kernels to be concurrently active on GPU, programmers need to downgrade the configuration they obtain by using the *Occupancy Calculator*. We refer to Kernel Concurrency (*KC*) as the maximum number of concurrently active kernels. The highest *KC* supported by CUDA as of compute capability 3.5 is 32. Due to the hardware resource limitations, a concurrency of X may be achieved by downgrading the configuration ($B$, $T$) to ($[B/X]$, $T$). We name such configuration *KC_X*.

For grid-level consolidation, at any given time there is only one active consolidated kernel that processes all the work from all threads in the parent kernel. Hence, we expect the best configuration to be the one that maximizes the device occupancy for a single kernel. Thus, we use *KC_1* as the default kernel configuration.

For block-level consolidation, each block in the parent kernel will spawn a consolidated kernel that will handle a smaller amount of work collected from a block in the parent kernel. We decide to downgrade the configuration by a factor of 16 and use *KC_16* as the default configuration.

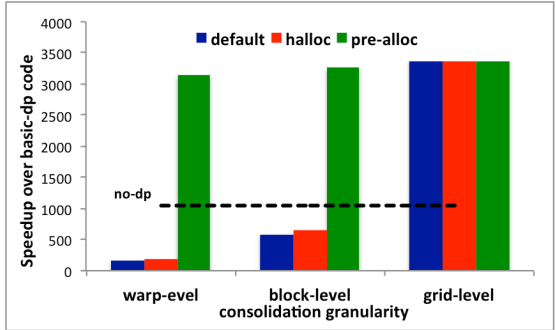

Figure 5. Performance of different buffer implementations (SSSP). All results are normalized to basic dynamic parallelism code.

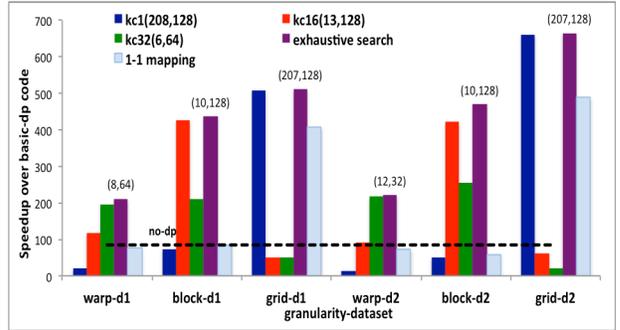

Figure 6. Performance of different kernel configurations (TD). All results are normalized to basic dynamic parallelism code.

For warp-level consolidation, each warp in the parent kernel will spawn a consolidated kernel that handles an even smaller amount of work collected from a warp, and a maximum concurrency of 32 can be easily achieved. Thus, we use *KC_32* as the default kernel configuration.

Because users may need to fine-tune the configuration for their applications to achieve the best performance, we also provide *#pragma* clauses that allow for user-specified kernel configurations.

## V. EXPERIMENTAL EVALUATION

In this section, we evaluate our proposed workload consolidation methods on various applications and datasets. Specifically, we compare the generated consolidated kernels with the original, basic DP kernels (*basic-dp* code template in Figure 1) and with flat GPU implementations of the considered algorithms (denoted by *np-dp*). In all the figures, *warp-level*, *block-level* and *grid-level* refer to the consolidation granularity considered.

We first evaluate the performance of using different memory allocators to implement the consolidation buffers. We then evaluate the effectiveness of our method to select the kernel configuration of nested kernels by comparing the resulting performance with the best performance achieved using the optimal kernel configuration found by exhaustive search. We then study the overall performance of the different consolidated kernels using the optimal allocator and kernel configuration. Finally, we use profiling to study the effect of our consolidation schemes on hardware utilization.

**Hardware and Software Setup**: We run all experiments on a workstation powered by two 6-core Xeon E5-2620 CPUs and a NVIDIA K20c GPU. We use CUDA runtime and compiler version 7.0. We use Nvidia Visual Profiler to collect the profiling data. We average the results of the experiments over multiple runs.

**Benchmark implementations**: The benchmarks used in our evaluations are *Single Source Shortest Path* (SSSP), *Sparse Matrix Vector Multiplication* (SpMV), *PageRank* (PR), *Graph Coloring* (GC), *Recursive Breadth-first Search* (BFS-Rec), *Tree Heights* (TH) and *Tree Descendants* (TD). Specifically, we consider the *basic-dp* and *no-dp*/flat implementations from [3, 5, 13, 14]. We use *basic-dp* implementations as baseline, and report the performance of flat kernels (*no-dp*) and consolidated kernels with different consolidation granularities. Since flat kernels achieve better performance than CPU implementations, we do not report results of CPU implementations.

**Datasets**: For applications based on graphs and sparse matrices, the datasets used are *CiteSeer* (used in SSSP, SPMV, PG) and *Kron_log16* (used in GC, BFS-Rec), both from the DIMACS challenges [15]. *CiteSeer* is a paper citation network with about 434 thousand nodes, 16 million edges and a node outdegree that varies from 1 to 1,199 across the graph (with an average value of 73.9). *Kron_log16* has 65 thousand nodes and 5 million edges, with a node outdegree that varies from 8 to 36,114. For the trees, we use datasets from [3]: *dataset1* is a depth-5 tree whose nodes have a number of children varying from 128 to 256 and only half of the non-leaf nodes have children; *dataset2* is a depth-5 tree whose nodes have a number of children varying from 32 to 128 and all non-leaf nodes have children.

### A. Implementation of the Consolidation Buffers

As explained in Section IV.E, the consolidation buffers can be implemented using three allocators: the default CUDA *malloc* allocator, the open-source high performance Halloc allocator [12] and our customized allocator. Figure 5 shows the performance results of workload consolidation on *SSSP* using these three allocators. In the figure, *default* refers to the CUDA *malloc/free* primitives, *halloc* to the Halloc allocator, and *pre-alloc* to our customized allocator. We can see that the *default* and *halloc* allocators achieve similar results in all cases. For block-level consolidation, the performance of both *default* and *halloc* are worse than that of the flat GPU code (*no-dp*), while *pre-alloc* achieves roughly 3x speedup over *no-dp*. This is due to the higher overhead introduced by *default* and *halloc* on every allocation operation. This overhead also contributes to a 5.7x performance gap between *pre-alloc* and *default*/*halloc* in case of block-level consolidation. The performance degradation of *default* and *halloc* is even worse (20x slowdown) for warp-level consolidated code, which requires more frequent buffer allocation operations. Since grid-level consolidation only requires a single buffer, in this case there is no obvious performance difference among these three allocators. In the remaining experiments, we only show the results reported using the better performing *pre-alloc* allocator.

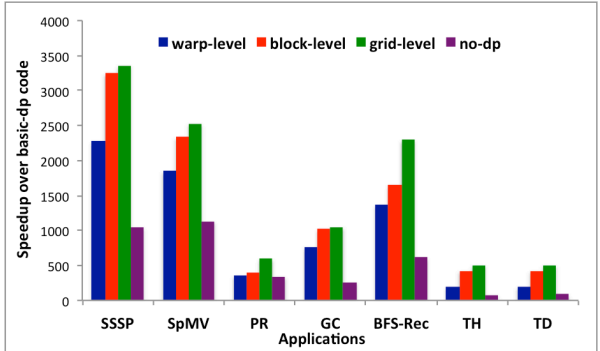

Figure 7. Overall speedup over basic dynamic parallelism.

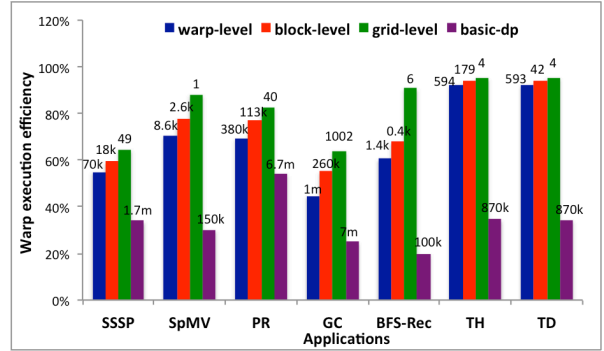

Figure 8. Warp execution efficiency across benchmarks.

*B. Selection of the Kernel Configuration*

In Section IV.E we discuss three configurations for consolidated kernels: KC_1, KC_16 and KC_32. These configurations allow the consolidated kernels to achieve concurrency levels (maximum number of concurrently active kernels) of 1, 16 and 32, respectively. We compare these configurations with two additional configurations schemes, *1-1 mapping* and *exhaustive search*. The *1-1 mapping* configuration indicates that the kernel is configured to have as many blocks (or threads, in the case of thread-mapped child kernels) as the number of items in the buffer. The *exhaustive search* configuration is the best configuration we find from exhaustively searching the configuration space [16]. In Figure 6 we report the results on tree descendants for all considered consolidation granularities over two datasets.

We first compare the three proposed configurations. KC_1 works best for grid-level consolidation; KC_16 works best for block-level consolidation; and KC_32 works best for warp-level consolidation. This meets our expectations and is coherent with the analysis presented in Section IV.E. We then compare KC_1 for grid-level, KC_16 for block-level and KC_32 for warp-level consolidation with *1-1 mapping*. As can be seen, our solution performs much better, especially for block- and warp-level consolidation. This is because the varying block size of *1-1 mapping* lowers the Kernel Concurrency and increases the size of the pending queue, leading to higher overhead and degraded performance. At last, we compare our scheme with the best configuration found by *exhaustive search*. We observe that the configurations selected by our method (KC_1 for grid-level, KC_16 for block-level and KC_32 for warp-level consolidation) achieve on average 97% of the performance of the optimal configuration found by exhaustive search.

The same experiments conducted on the other benchmarks using various datasets report similar results. In conclusion, our configuration selection method for nested kernels is effective and leads to nearly optimal performance. In all the remaining experiments, we use KC_1, KC_16 and KC_32 configurations for kernels consolidated at the grid-level, block-level and warp-level granularities, respectively.

*C. Overall Performance*

Figure 7 presents the overall speedup of kernels consolidated at different granularities over *basic-dp*. The chart also reports the speedup achieved by flat parallel code (*no-dp*) over the baseline. As can be seen, the *basic-dp* implementation suffers from severe performance degradation due to the significant kernel management overhead and the limited GPU utilization. Even compared with the flat GPU kernels, *basic-dp* reports slowdown factors from 80 to 1100. Warp-level consolidation improves the performance of *basic-dp* on average by a factor of 1000x but in some cases is not significantly better than the flat GPU kernel (*no-dp*). Block-level consolidation outperforms warp-level consolidation, and grid-level consolidation achieves the best performance across all benchmarks. Even if warp-level consolidation has the benefit of very low synchronization overhead, when compared to block- and grid-level code, it suffers from the more significant overhead introduced by the additional child kernel launches. Grid-level consolidation reduces the number of child kernel launches dramatically, and thus achieves the best performance despite its extra synchronization overhead. On average, warp-level, block-level and grid-level consolidation outperform *basic-dp* by a factor of 999, 1357 and 1459, respectively, and *no-dp* by a factor of 2.18, 3.26 and 3.78, respectively.

*D. Profiling Results*

In this section, we analyze the improvements in hardware utilization achieved by workload consolidations.

Figure 8 shows the overall *warp execution efficiency*, which is defined in the CUDA documentation [17] as "the ratio of the average active threads per warp to the maximum number threads per warp". For each application, we show the results reported by the *basic-dp* implementation and the three considered workload consolidation schemes. On top of the bars we report the number of child kernel invocations performed in each case. First, we can observe that the proposed consolidation methods reduce the number of kernel invocations to 0.07%-14.48% of the ones performed by the corresponding *basic-dp* code. For instance, in *PageRank*, consolidation reduces the number of kernel invocations from 6.7 million (*basic-dp*) to 380 thousand (*warp-level*), 113 thousand (*block-level*) and 40 (*grid-level*). Second, average warp execution efficiencies are improved from 33.2% (*basic-dp*) to 69.3% (*warp-level*), 75% (*block-level*) and 83.1% (*grid-level*). The warp execution efficiency measured by Nvidia profiler includes not only parent and child kernels execution, but also child kernel launch overhead. Child kernel launches will take more clock cycles than buffer insertion operations, decreasing the warp efficiency. Consolidation replaces kernel launches with buffer insertions; as a result, it reduces the overhead of warp divergence and leads to improved overall warp efficiency.

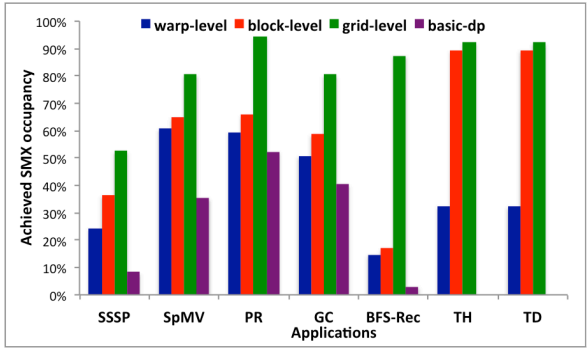 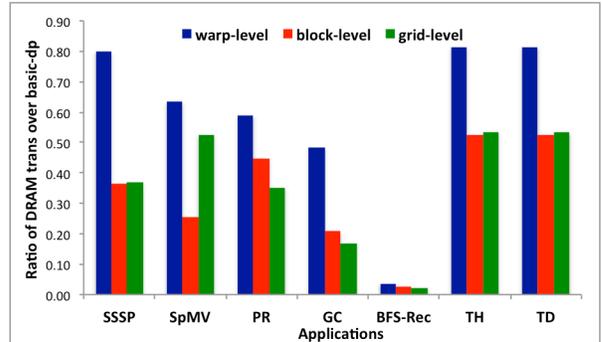

Figure 9. SMX occupancy (achieved hardware utilization). For TH and TD, the profiler reports "overflow" errors in basic-dp.

Figure 10. DRAM transactions ratio over basic dynamic parallelism.

Figure 9 shows the achieved streaming multiprocessor occupancy, which measures the ratio of average active warps over maximum warps supported per streaming multiprocessor [17]. On average, workload consolidation improves this metric from 27.9% (*basic-dp*) to 39.3% (*warp-level*), 60.3% (*block-level*), and 82.9% (*grid-level*). Recall that in *basic-dp* each thread launches a "small" kernel. Hence, the GPU device is filled with many such "small" concurrent kernels. As mentioned in Section IV.B, there is a hardware limitation on the maximum number of concurrent kernels that a GPU can accommodate. On K20c, this limit is 32. Therefore, in the *basic-dp* case, thirty-two "small" concurrent kernels will underutilize the hardware. Workload consolidation, on the other hand, increases the average child kernel size and improves resource utilization.

To measure the efficiency of memory accesses, we monitor the numbers of DRAM transactions (read+write) performed by each kernel. Figure 10 shows the ratio between the number of DRAM transactions performed by each consolidated kernel and those performed by the *basic-dp* code. In all cases, the total DRAM transactions are reduced. Specifically, *warp-level*, *block-level* and *grid-level* consolidation lead to 60%, 34% and 36% of the original transactions, respectively. This reduction in memory transactions can be motivated as follows: first, the consolidation increases the average child kernel size, thus leading to better caching behavior and memory bandwidth utilization; second, a decrease in the number of nested kernels will lower the chance of swapping parent kernels out, therefore reducing the memory transaction overhead associated to kernel swapping; third, the decreased number of nested kernels reduces the use of the virtualized pool within the pending queue, lowering the overhead of virtual pool management. It can also be noticed that, for some benchmarks (e.g. *SpMV*), block-level achieves better memory utilization than grid-level consolidation. This is due to the overhead associated to our global synchronization mechanism.

## VI. RELATED WORK

In recent years there has been an increasing interest in the study of the effective use of GPUs for applications with complex or data dependent parallelism. These studies have started well before the introduction of DP on GPUs. Burtscher et al [18, 19], Shuai et al [20], and Wu et al. [21] have conducted performance and power characterizations of applications with runtime parallelism on GPUs. In the last few years, many such applications have been successfully accelerated on GPUs [5, 22-28]. Other work has targeted the effective parallelization of applications with nested data parallelism on architectures with flat hardware parallelism. Blelloch and Sabot have proposed *flattening* techniques [8] and NESL language [9] to vectorize programs with nested data parallelism. Bergstrom and Reppy [29] ported NESL to GPUs. Other studies have focused on mechanisms to facilitate the deployment of applications with runtime parallelism on GPUs. For example, Gupta et al. [30] have introduced the persistent threads method to process dynamic tasks. Similarly to these works, we also consider applications with runtime and nested parallelism. However, we focus on the effective use of DP for these computations.

Although the effectiveness of dynamic parallelism has been demonstrated on certain applications (such as clustering algorithms [31], computation of the Mandelbrot set [1] and a particle physics simulation [32]), it has been shown that, because of the non-negligible overhead of this feature, the naïve use of DP can actually slow down the performance [2-4]. Wang et al. [2] have performed a characterization of DP-based implementations of unstructured applications, focusing on the analysis of their control flow behavior, their memory access patterns and the DP overhead. Yang and Zhou [4] have proposed a compiler framework to support nested thread-level parallelism without using DP. Their solution, which leads to spawning a massive number of threads, does not apply to recursive computations and is less effective on applications that exhibit high degrees of thread-level parallelism even before the proposed code transformations. Chen and Shen [33] have proposed a child kernel removal method based on the reuse of parent threads. Unlike ours, their method does not apply to recursive computations. In our previous work [3], we have proposed code parallelization templates – with and without DP – to facilitate the efficient execution of nested parallel codes on GPUs. While the proposed DP-based templates for irregular loops are similar to the codes generated by block- and grid-level consolidation, the DP-based templates for recursive computations are fundamentally different and apply only to hierarchical data structures, such as trees and graphs. In general, the parallelization templates presented in [3] are not uniform across the two considered computational patterns. In this paper, we have proposed generic code transformation techniques that apply equally to irregular loops and recursive applications, and we have automated these code transformations through compiler integration. Wang et al. [34] have proposed a hardware architecture to support lightweight block execution of dynamically launched kernels. Conversely, our method is purely software-based, and is thus applicable on any GPU that supports dynamic parallelism, requiring no modification to the architecture.

## VII. CONCLUSION

In this paper, we have proposed a workload consolidation mechanism to improve the efficiency of DP-based codes. Our code transformation technique essentially aggregates child kernels dynamically spawned by different threads into a single kernel launch. Our method can significantly reduce the overhead of managing dynamic kernels and increase the level of concurrency of DP-based codes. We automate the proposed code transformations by implementing them in a directive-based compiler, thus allowing programmers to write simple and unoptimized DP-based codes focusing on functionality rather than worrying about code efficiency and performance. Our experimental evaluation conducted on algorithms exhibiting irregular loops and parallel recursion shows that kernels consolidated with the proposed code transformations achieve an average speedup in the order of 1500x over basic implementations using DP and an average speedup of 3.9x over optimized flat GPU kernels.

ACKNOWLEDGEMENT

This work has been supported by NSF awards CNS-1216756 and CCF-1452454 and by a gift from NEC Laboratories America and equipment donations from Nvidia Corporation.